\documentclass[aps,pre,twocolumn,showpacs,
               preprintnumbers,amsmath,amssymb,floatfix]{revtex4}

\usepackage[dvips]{graphicx}
\usepackage{bm}
\usepackage{amssymb}

\newcommand{\bd}{../../../bib}

\newcommand{\emath}[1]{\ensuremath{#1}}

\newcommand{\Vect}[1]{\textbf{#1}}  
\newcommand{\Vectm}[1]{\bm{#1}}  
\newcommand{\Tensor}[1]{\textbf{\underline {#1}}}
\newcommand{\derivative}[3] {\emath{\displaystyle \frac{#1#2}{#1#3}}}  
  
\newcommand{\Grad}[1]{\Vectm{\nabla}\!{#1}}

\newcommand{\Div}[1] {\Vectm{\nabla} \cdot {#1}}

\newcommand{\fig}{\begin{figure}}
\newcommand{\efig}{\end{figure}}
\newcommand{\figfull}{\begin{figure*}}
\newcommand{\efigfull}{\end{figure*}}

\newcommand{\figref}[1]{Fig. \ref{#1}}

\newcommand{\subfigsref}[3]{Fig. \ref{#1}(#2--#3)}

\newcommand{\subfigref}[2]{Fig. \ref{#1}(#2)}
\newcommand{\Subfigref}[2]{Figure \ref{#1}(#2)}
\newcommand{\Table}{\begin{table}}  
\newcommand{\etable}{\end{table}}  
\newcommand{\eq}{\begin{equation}}  
\newcommand{\eeq}{\end{equation}}  
\newcommand{\eqa}{\begin{eqnarray}}  
\newcommand{\eeqa}{\end{eqnarray}}  

\newcommand{\secref}[1]{(\S\,\ref{#1})}

\newcommand{\Secref}[1]{Section \ref{#1}}

\begin{document}

\preprint{rev1.03}

\title{New boundary conditions for granular fluids}
\author{M. D. Shattuck}
\email[]{shattuck@ccny.cuny.edu}
\homepage{http://gibbs.engr.ccny.cuny.edu}
\affiliation{Benjamin Levich Institute and Physics Department,
The City College of the City University of New York
140th and Convent Ave., New York, NY  10031}

\date{\today}

\begin{abstract}

We present experimental evidence, which contradicts the the standard
boundary conditions used in continuum theories of non-cohesive
granular flows for the velocity normal to a boundary
$\Vect{u}\cdot\hat{n}=0$, where $\hat{n}$ points into the fluid.  We
propose and experimentally verify a new boundary condition for
$\Vect{u}\cdot\hat{n}$, based on the observation that the boundary
cannot exert a tension force $\Vect{F}_b$ on the fluid.  The new boundary
condition is $\Vect{u}\cdot\hat{n}=0$ if $\Vect{F}_b\cdot\hat{n}\ge 0$ else
$\hat{n}\cdot\tensor{P}\cdot\hat{n}=0$, where $\Tensor{P}$ is the
pressure tensor.  This is the analog of cavitation in ordinary fluids,
but due the lack of attractive forces and dissipation it occurs
frequently in granular flows.

\end{abstract}



\maketitle

\section{Introduction} 

Granular materials, collections of particles which dissipate energy
through inter-particle interactions have tremendous technological
importance and numerous applications to natural systems.  They also
represent a serious challenge to statistical and continuum modeling,
due to the small number of particles and dissipation.  A fundamental
understanding of granular systems, comparable to our current
understanding of fluids and solids, does not exist today
\cite{kadanoff99} but would have far reaching implications across many
industries.

The idealized granular system, a collection of identical {\em
inelastic} hard spheres, is a laboratory scale analog of the canonical
system in kinetic theory --- the {\em elastic} hard sphere gas.  Using
this analogy, an {\em inelastic} version of the Boltzmann-Enskog
equation have be derived {\cite{chapman70,jenkins85}} as well as an
{\em inelastic} version of continuum equations for mass, momentum, and
energy balance
{\cite{chapman70,jenkins85,goldshtein95,sela96,brey96,santos98,vannoije98b}}
nearly identical to the Navier-Stokes equations.  These type of
equations can produce a quantitative description of dilute granular
flows with an accuracy of 1\% \cite{rericha2002}.  Boundary conditions
for velocity and granular temperature have been derived to complete
these equation \cite{jenkins91,richman92,jenkins92,jenkins97}, which
assume that the fluid never leaves the boundary. 

\fig
\includegraphics[width=3.375in]{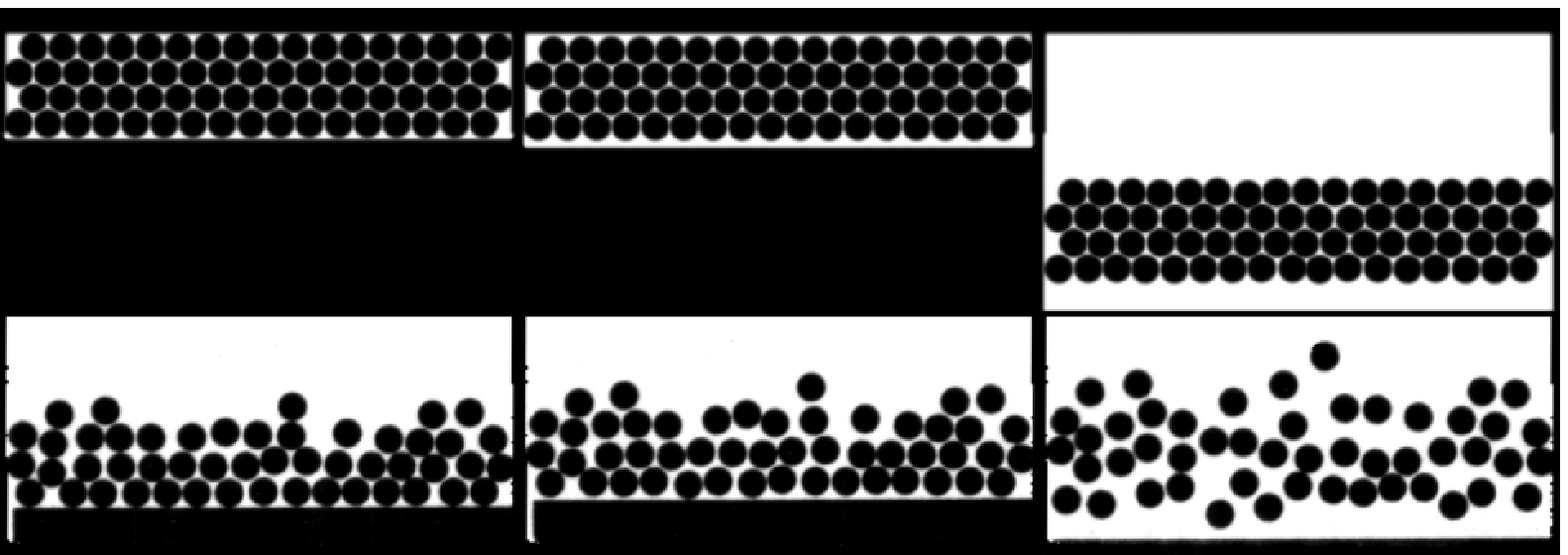}
\caption{photographs of a 2D granular layer under free fall conditions
(top row) and vibrated (bottom row).  The layer is initially in
contact with the boundary (first column), but then moves away (second
and third column). }
\label{cell}
\efig

Normal fluids remain attached to boundaries due to two effects ---
adhesion and pressure (external and internal). External pressure is
unimportant and adhesion is not present in non-cohesive granular
materials leaving the internal pressure alone to hold a granular fluid
to a boundary. However, if the force on the granular fluid accelerates
the fluid to a velocity greater than the average root-mean-squared
particle velocity (square root of the granular temperature), then the
boundary and the granular fluid will separate.  This is the analog of
cavitation in normal fluids.  While cavitation is unusual in normal
fluids, low granular temperature at boundaries due to inelastic loss
combined with the lack of external pressure and adhesion makes
cavitation common in granular fluids (see \figref{cell}).  Common
examples include rotating drums and vibrated layers.  Many features of
pattern formation in vibrated layers \cite{melo94} can be understood
in terms of the time that the layer spends off of the plate
\cite{melo95}.  It difficult to explain this dependence with models
which do not allow the granular fluid to leave the boundary.

\section{New Boundary Condition} 

The standard boundary condition
\cite{jenkins91,richman92,jenkins92,jenkins97} for the velocity normal
to the wall is 
\eq
(\Vect{u}-\Vect{v})\cdot\hat{n}=0,
\label{eq:obc}
\eeq 
for the fluid velocity \Vect{u} on a boundary with inward unit
normal $\hat{n}$ and wall velocity \Vect{v}.  In order to enforce this
condition the boundary must exert a force $\Vect{F}_b$ on the granular
fluid.  Because there is no attractive force between the grains and
the boundary $\Vect{F}_b\cdot\hat{n}\ge 0$.  When the force needed to
maintain zero velocity is negative the boundary condition must be
changed.  Since the boundary force then must be zero a no stress
condition $\hat{n}\cdot\Tensor{P}\cdot{n}=0$ is the logical choice.
As the normal wall velocity will no longer be zero we must have a
another condition to enforce the no particle flux condition at the
boundary.  Analytically more work needs to be done to understand this
new condition on the flux $\Phi=\int \rho\Vect{u}\cdot\;d\Vect{S}$.
Numerically, we treat the density and velocity as centered in the grid
and the flux is defined on the boundaries of the grid.  In that case
the density and velocity in the cell nearest the boundary are both
non-zero, but the flux on the wall is zero (see \secref{sim} for
details of the flux condition). 
Altogether we have
\eq \left\{
\begin{array}{lll}
(\Vect{u}-\Vect{v})\cdot\hat{n}=0  & \text{if}  & \Vect{F}_b\cdot\hat{n}> 0\\
\hat{n}\cdot\Tensor{P}\cdot{n}=0, \Phi=0  & \text{else} &
\Vect{F}_b\cdot\hat{n}\le 0.
\end{array}\right.
\label{eq:nbc}
\eeq

\fig
\includegraphics[width=3.375in]{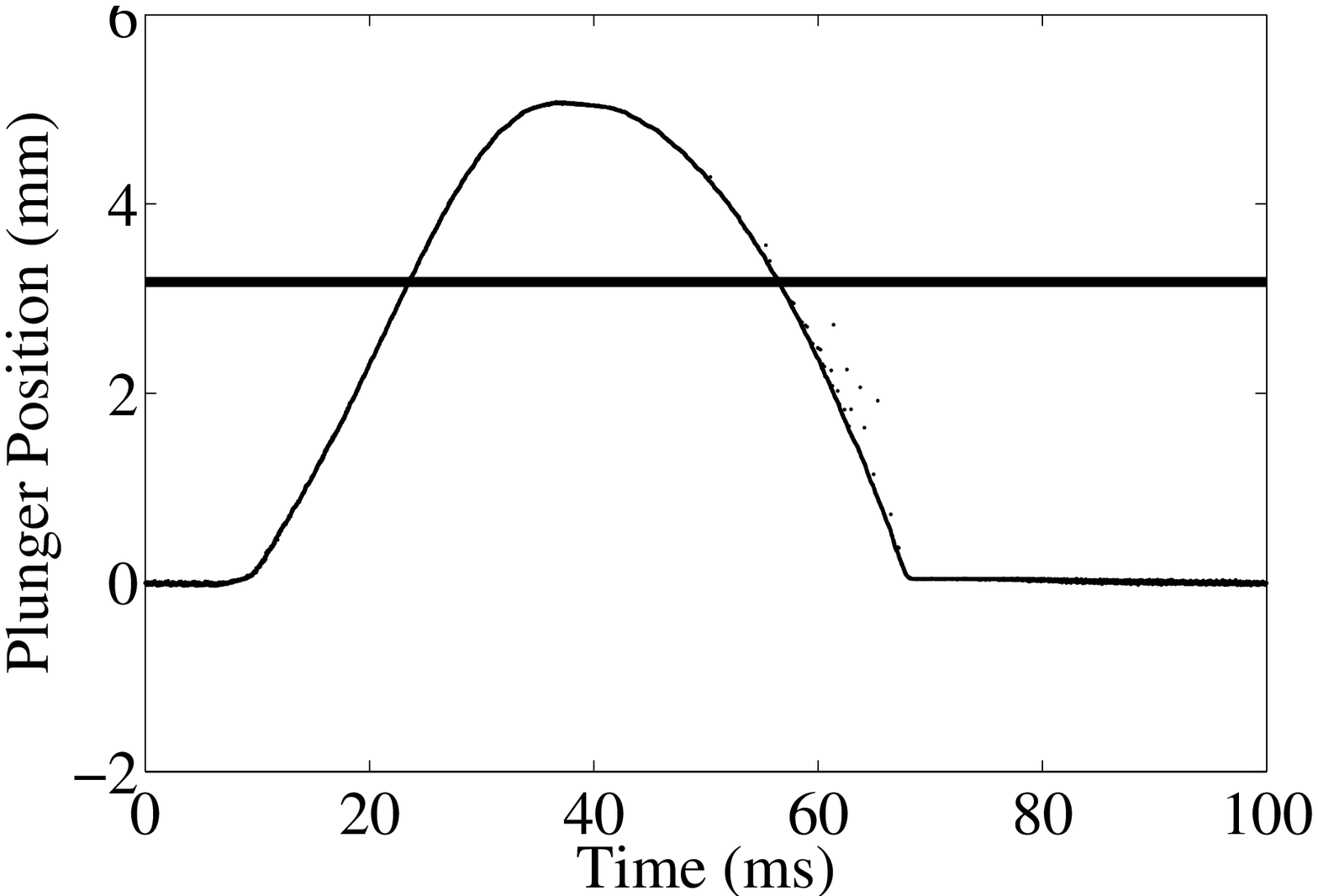}
\caption{Plot showing the lower boundary of the vibrated system as a
  function of time.  94 cycles at 10Hz are overlaid.  The horizontal
  line shows the diameter of a particle.}
\label{plate}
\efig

\section{Experiment} 

We have performed two types of experiments to elucidate the role of
boundary conditions in continuum equations of motion for granular
materials --- free-fall and vibrated. We place $N$ ($51$--$117$)
spherical stainless steel ball bearings of diameter $D=3.175$ mm in a
container 17.5 $D$ wide by 20 $D$ tall by 1 $D$ deep as shown in
\figref{cell}.  We define the number of rows $R=N/17$, where 17 is the
number of particles to fill an entire row in a close-packed structure.
We control a thin plunger, which slides through a slot in the bottom
of the cell.  For the free-fall experiment [\subfigref{cell}{top row}]
the plunger pushes all of the particle to the top of the cell to start
the experiment.  The particles are then vibrated in order to compact
them into a perfect hexagonal packing [\subfigref{cell}{top-left}].
Then the plunger is pulled downward with an initial acceleration of
$4g$, where $g$ is the acceleration of gravity.  The particles are
then free to fall under gravity.  This process is under computer
control and can be repeated many times.  In the vibrated experiments
[\subfigref{cell}{bottom row}] the plunger oscillates sinusoidally,
but is offset from the bottom of the container to produce a
half-sine-wave excitation.  The measured position of the plate is
shown in \figref{plate}.  94 oscillations are superimposed to show the
repeatability of the drive.  The excitation is characterized by the
non-dimensional maximum acceleration $\Gamma = A(2\pi f)^2/g$, where
$A$ is the maximum amplitude of the plunger and $f$ is the frequency.
Using high-speed digital photography we measure the positions of the
plunger and all of the particles in the cell with a relative accuracy
of 0.2\% of $D$ or approximately $6 \mu$m at a rate of 840 Hz. We
track the particles from frame to frame and assign a velocity to each
one, typically $\sim D/5$ per frame.

\fig
\includegraphics[width=.95\columnwidth, bb=69 0 471 359, clip=true]{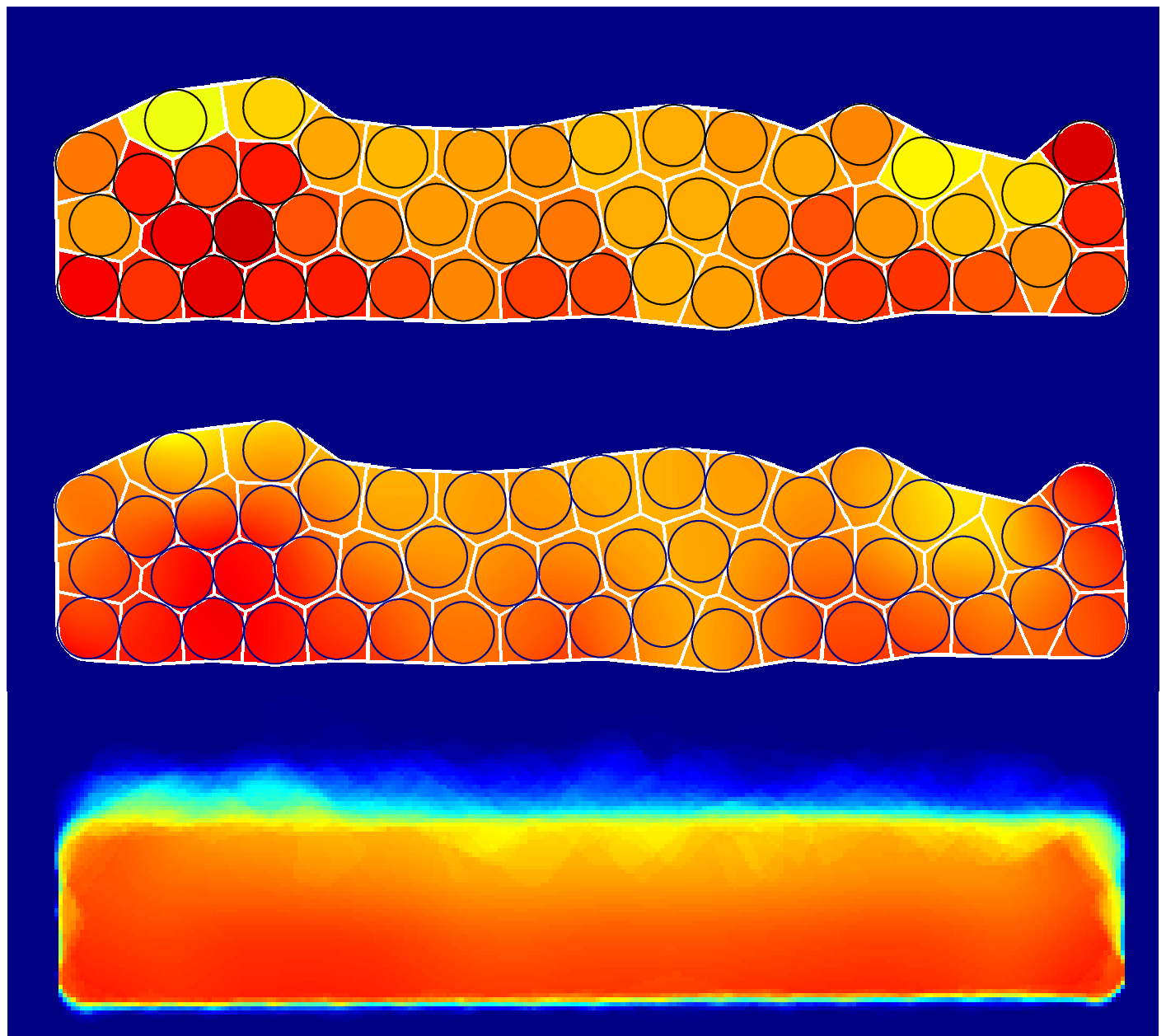}
\caption{Color online: Steps in the spatial and temporal averaging
  process for density field. (top) Modified Voronoi cells, 
(middle) footprint preserving spatial average over 2 particle
diameters, (bottom) average over phase (time).}

\label{voronoi}
\efig

To compare with the continuum theories through our simulations (see
\Secref{sim}), we must average the particle trajectories to created
density, velocity, and temperature fields.  We are interested in
studying the flow as the grains separate from the boundary so we focus
on the density fields.  To create average fields we run each flow
condition 90-100 times.  In each frame, we construct a modified
Voronoi cell around each particle and fill it with a uniform number
density of 1 particle divided by the area of the cell.  We modify the
Voronoi cells on the border by including only the area that overlaps
with a sector created by two lines both tangent to the particles edge
and tangent to two other neighbors less than two particle diameters
away.  This is the union of the convex hull formed by all all particle
pairs whose centers lie within 2 particle diameters of each other.
The results of such a construction are shown in
\subfigref{voronoi}{top}. Next, we apply a mask preserving spatial
average over 2 particle diameters [\subfigref{voronoi}{middle}].  The
mask is formed by the union of all of the modified voronoi cells.
Finally, we bin the frames according the phase of the cycle and
average each bin [\subfigref{voronoi}{bottom}].  From these images we
have determined that the flow is uniform in the horizontal $x$
direction for both free-fall and vibrations.  Therefore, $x$ to obtain
one dimension density field along the vertical $z$ direction.
Space-time intensity plots of these densities are shown in
\figref{spacetime}.

\figfull
\includegraphics[width=\textwidth, bb=20 0 524 138,clip=true]{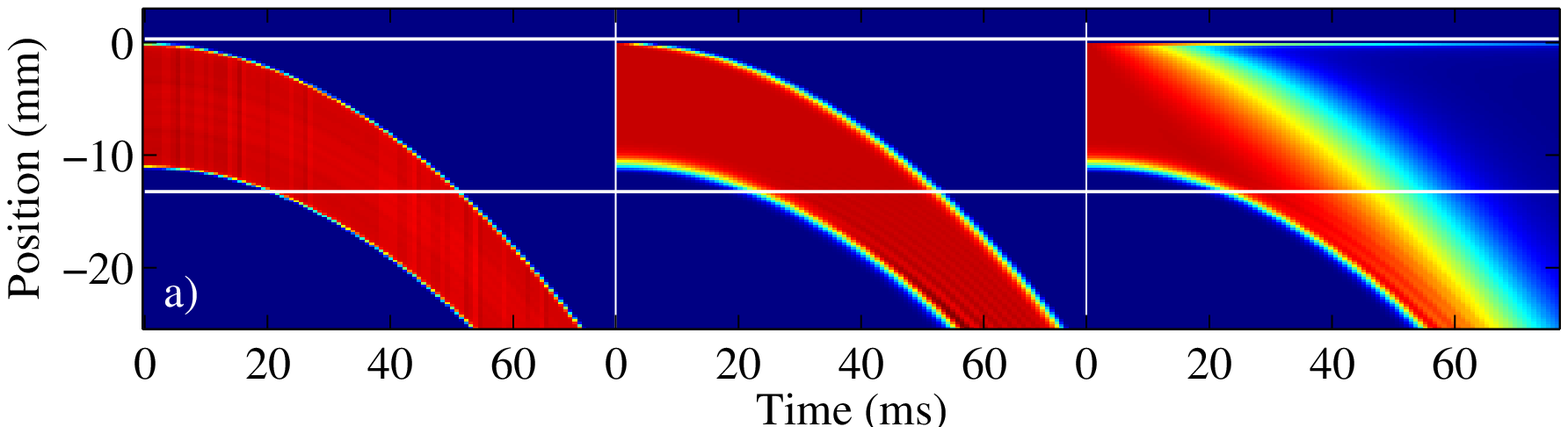}
\includegraphics[width=\textwidth, bb=20 0 524 138,clip=true]{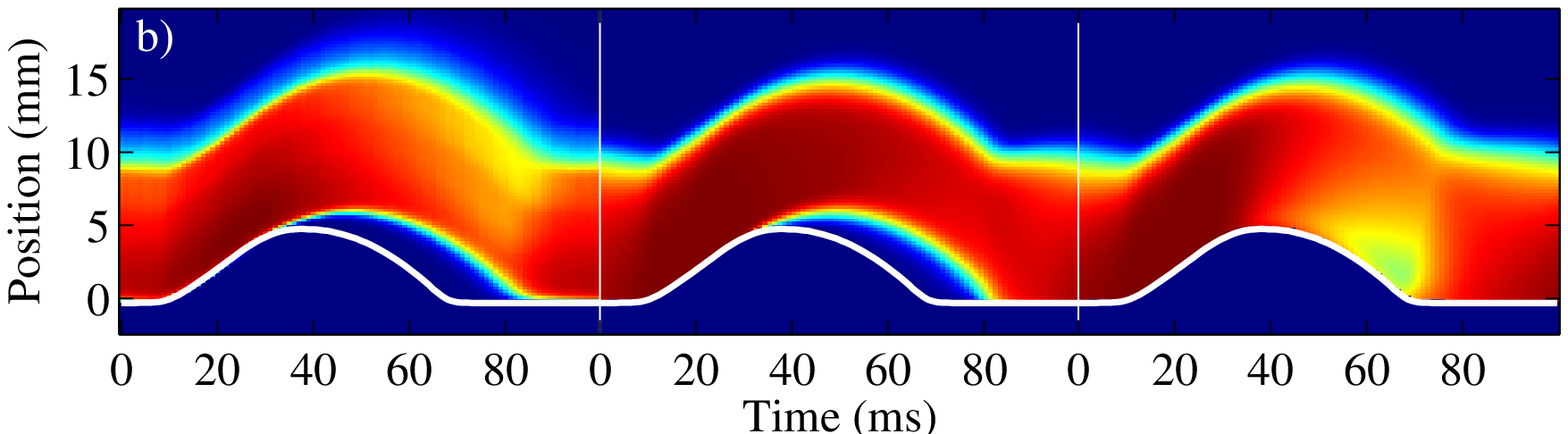}
\caption{Color online: Space-time plots of the volume fraction for a)
  free-fall and b) vibrated.  Experiments are shown in the first column,
  simulations with the new boundary conditions in the the second
  column, and simulations with the old boundary conditions in the third
  column.  In a) the upper white line is the top of the cell.  The
  lower line show where the number flux is calculated in
  \subfigref{compare}{a}.  In b) the line is the position of the bottom
  plunger.  The separation of the fluid from the plate can clearly be
  seen in the experiment and the simulation using the new boundary
  conditions.}
\label{spacetime}
\efigfull

As a further quantitative comparison, we calculated the discharge rate
for the free-fall case [\subfigref{compare}{a}] and the acceleration
of the center of mass $\Vect{a}_{\text{CM}}$ minus $\Vect{g}$ for all
cases [\subfigsref{compare}{b}{d}].
$M(\Vect{a}_{\text{CM}}-\Vect{g})=\Vect{F}_b$, where $M$ is the total
mass of the fluid.  In \subfigsref{compare}{b}{d} we plot
$a_b=1/M\Vect{F}_b\cdot\hat{z}=(\Vect{a}_{\text{CM}}-\Vect{g})\cdot\hat{z}$
for both the free fall and the vibrating case.

\figfull
\includegraphics[width=.45\textwidth]{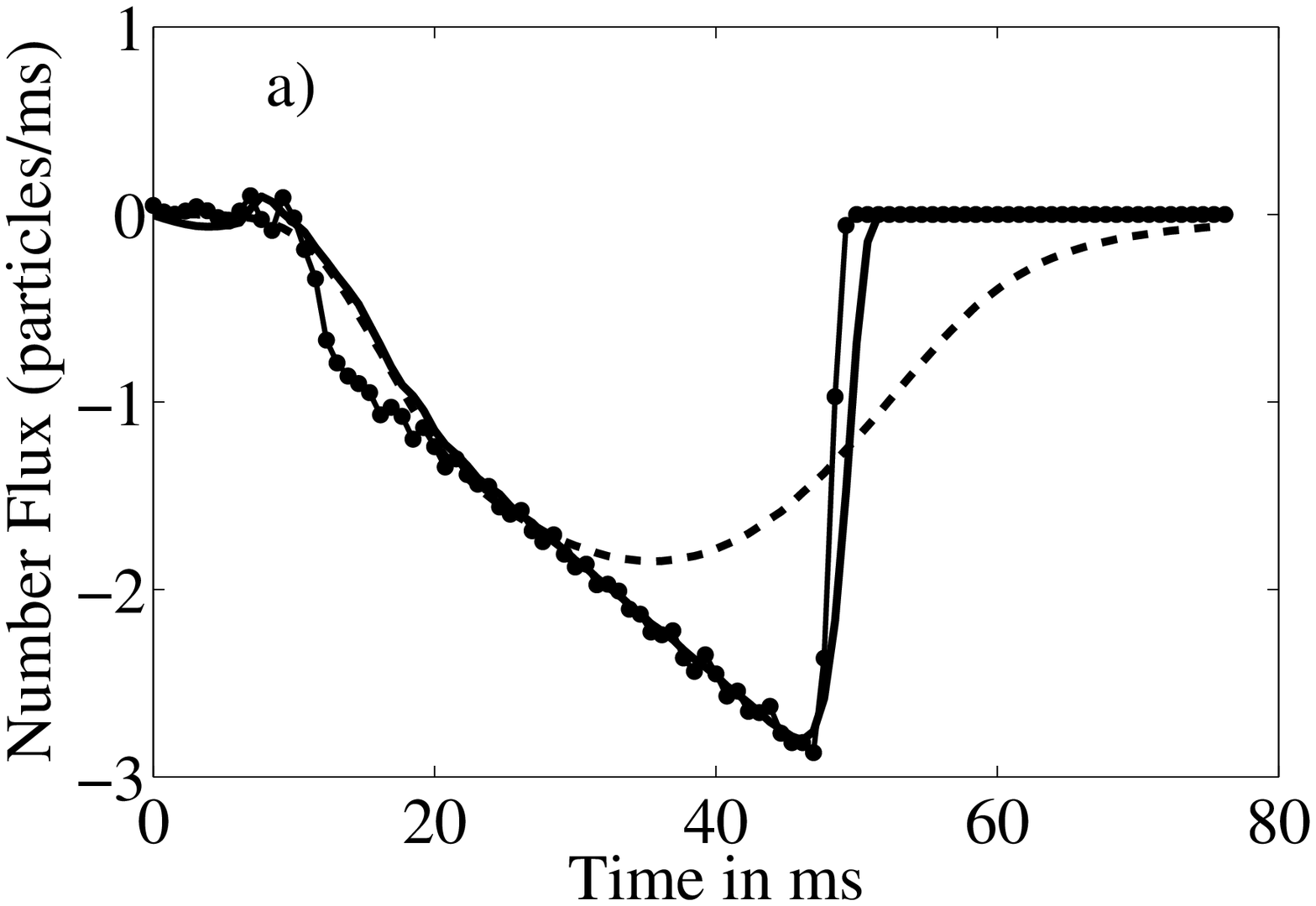}
\includegraphics[width=.45\textwidth]{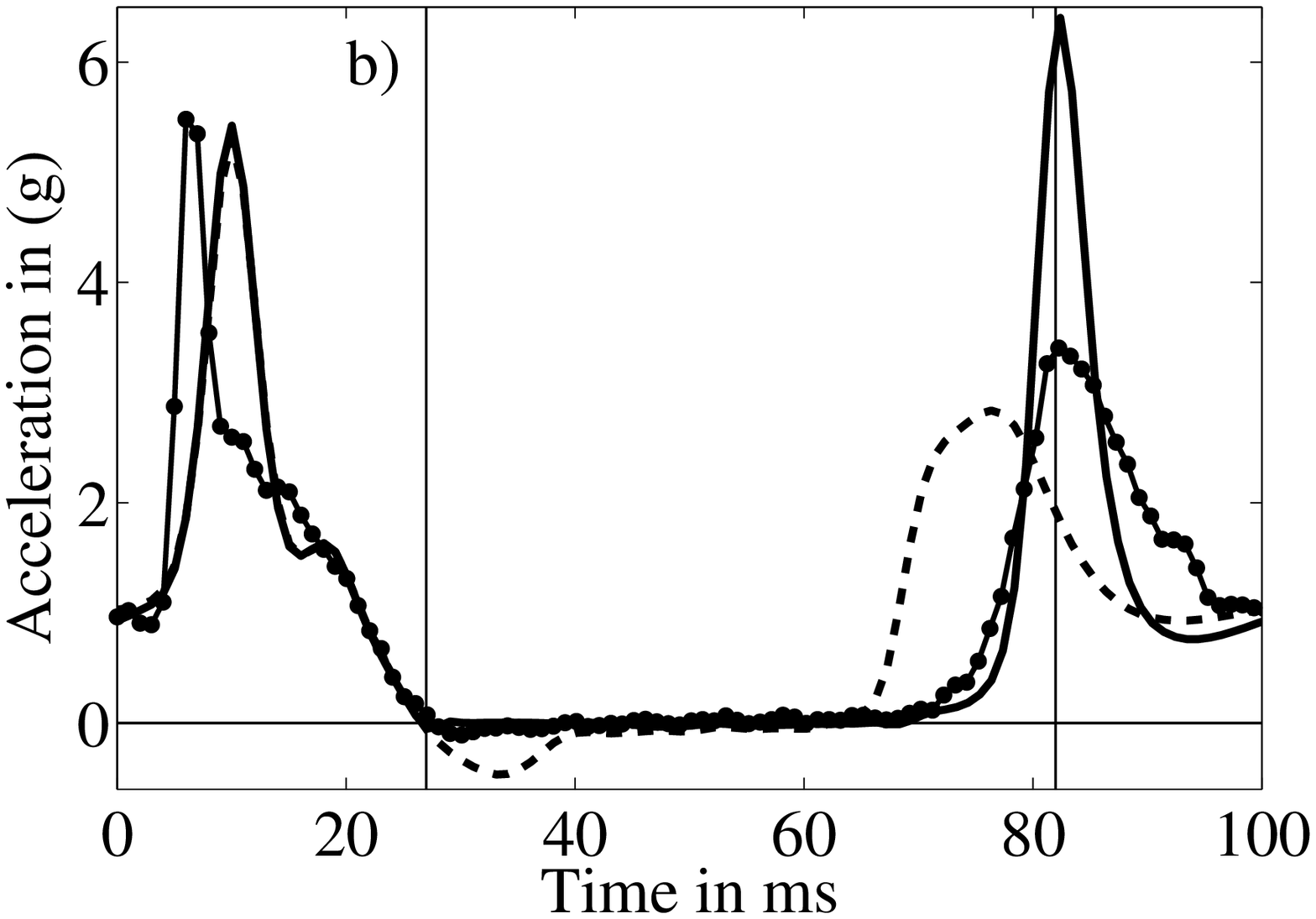}
\includegraphics[width=.45\textwidth]{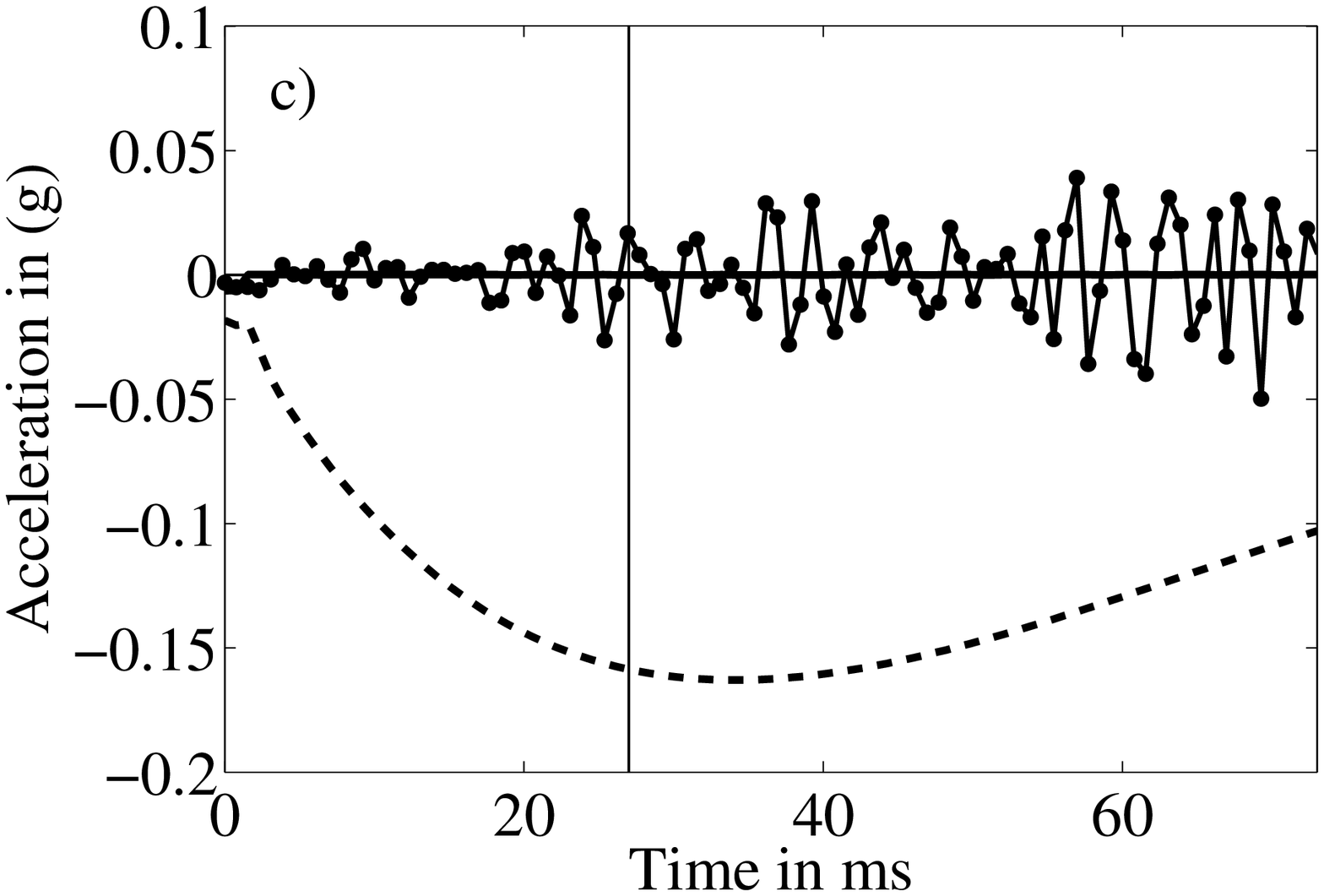}
\includegraphics[width=.45\textwidth]{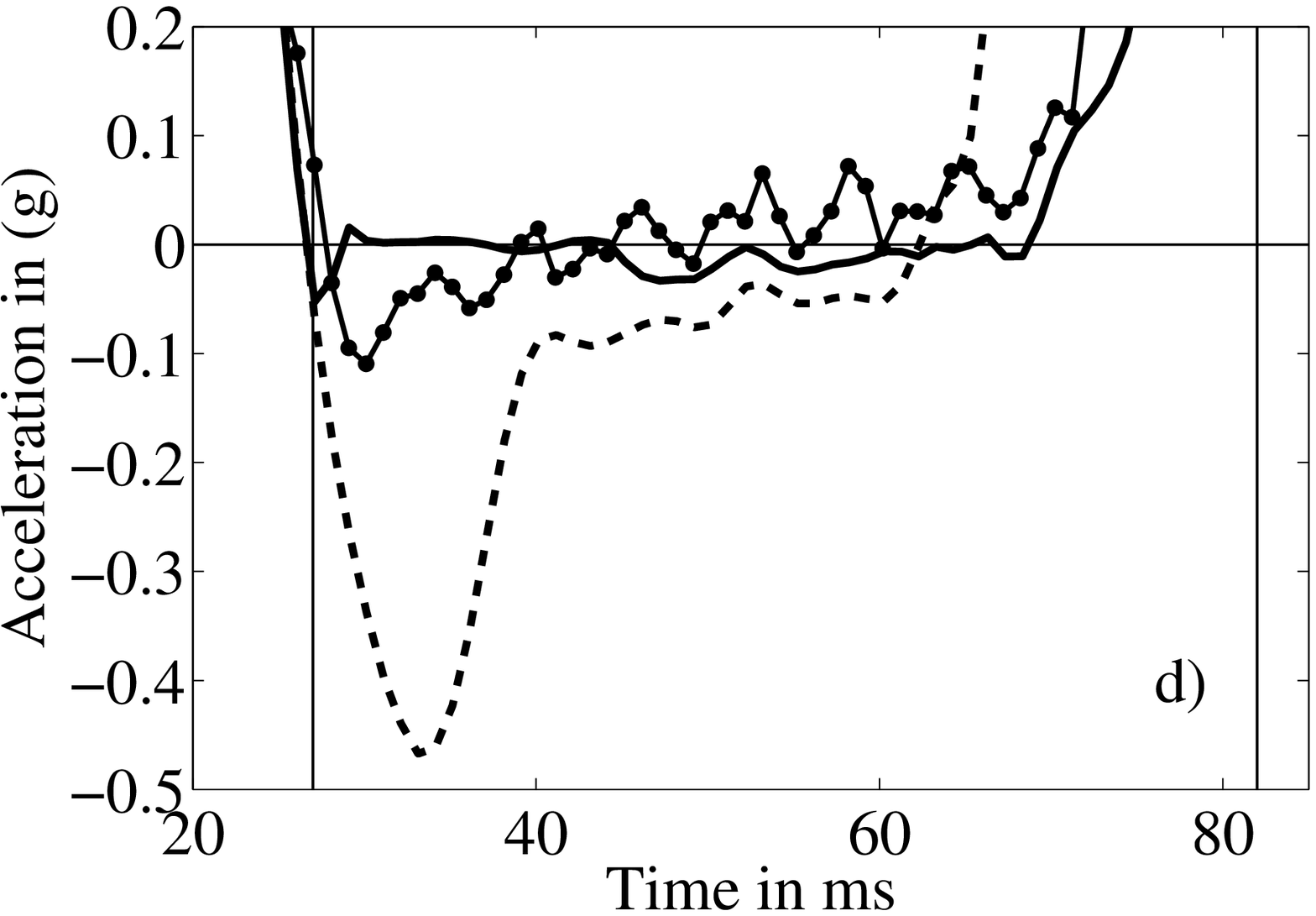}
\caption{Comparison of experiment (connected points) and simulation
  with new boundary conditions (solid) and old boundary conditions
  (dashed) for free-fall (a,c) and vibration (b,d).  Plots of a)
  Number flux of particles passing the lower horizontal line in
  \subfigref{spacetime}{a} as a function of time and b-d) acceleration
  of the center of mass of the granular fluid.}
\label{compare}
\efigfull

\section{Simulation}\label{sim} 

We numerically integrate the continuum equations for the balance of
mass, momentum, and energy have been derived for two-dimensional (2D)
granular flows~\cite{jenkins85,jenkins85a}:
\eqa
\derivative{\partial}{\rho}{t} + \Div{(\rho\Vect{u})}&=&0\\
\rho(\derivative{\partial}{\Vect{u}}{t} +
\Vect{u}\cdot\Grad{\Vect{u}}) &=& -\Div{\Tensor{P}} -\Vect{g}\\
\rho(\derivative{\partial}{T}{t} +  \Vect{u}\cdot\Grad{T}) &=&
-\Div{\Vect{q}} - \Tensor{P}:\Tensor{E}-\gamma,
\label{balance}
\eeqa 
where $\rho$ is the mass density, \Vect{g} is the gravity
vector, $T$ is the granular temperature, \Vect{q} is the heat flux
vector, $E_{ij}={\frac{1}{2}}(\partial_i u_j + \partial_j u_i)$ are
the elements of the symmetrized velocity gradient tensor \Tensor{E},
and $\gamma$ is the temperature loss rate. The constitutive relations
for the pressure tensor $\Tensor{P}$ and heat flux $\Vect{q}$
are \eq {\bf \underline P} = (P - 2 \lambda {\rm Tr}{\bf \underline
E}){\bf \underline I} - 2 \mu ({\bf \underline E} - ({\rm Tr}{\bf
\underline E}){\bf \underline I})
\label{newtonslaw}
\eeq
and
\eq
{\bf q} = - \kappa {\bf \nabla} T,
\label{fourierslaw}
\eeq
where Tr denotes trace and  ${\bf \underline I}$ is the unit tensor.
The 2D equations close~\cite{jenkins85a} with the equation of state, which
is the ideal gas equation of state with a term that includes dense gas
and inelastic effects,
\eq
P = \rho T [1 + (1+e) G(\nu)],
\label{state}
\eeq 
where $e$ is the coefficient of restitution,
$\nu=\frac{\rho\pi\sigma^2}{4m}$ is the volume fraction, $\sigma$ is
the diameter of the particles, $m$ is the mass of the particles, and
$G(\nu)=\nu g(\nu,\sigma)$, where $g(\nu,\sigma)$ is the value of the
radial distribution function at a distance of one particle diameter.
We use a temperature dependent 
\eq 
e(T)=
\left\{
\begin{array}{ll}
  1-(1-e_0)\sqrt{T/T_0}^{1/5} & if T<T_0\\
  e_0 & otherwise.
\end{array}\right.
\eeq
This mimics the experimental evidence that $e_0$ goes to zero for
small velocities.  We use a form for $G(\nu)$ developed by Torquato
\cite{torquato95}, which is an analytical fit to molecular dynamics
simulation at high $\nu$ and the Carnahan and Starling
\cite{carnahan69} geometric series approximation to the first few
viral coefficients at low $\nu$.  We have changed the functional form
of $G(\nu)$ and the details of the flow change, but not the
qualitative behavior.  The bulk viscosity,
\eq 
\lambda = 2\rho\sigma\sqrt{\frac{T}{\pi}}G(\nu),
\eeq 
the shear viscosity, \eq \mu = {\frac{\rho\sigma\sqrt{\pi T}}{8}}
[{\frac{1}{G(\nu)}} + 2 + (1+\frac{8}{\pi})G(\nu)],
\label{eq:mu}
\eeq
the thermal conductivity,
\eq
\kappa = {\frac{\rho\sigma\sqrt{\pi T}}{2}} [\frac{1}{G(\nu)} + 3 + (\frac{9}{4}+\frac{4}{\pi})G(\nu)],
\label{eq:kappa}
\eeq
and the temperature loss rate per unit volume,
\eq
\gamma=\frac{4\rho G(\nu) T^{3/2}}{\sqrt{\pi}\sigma}(1-e^2).
\label{gamma0eq}
\eeq 

We use a second-order-space adaptive first-order-time
finite-differencing scheme to integrate these equation.  We use
off-center-differencing at the boundaries to maintain second-order
accuracy.  The code uses an adaptive time step based on a modified
Courant condition combined with an empirically determined maximum and
minimum time step.  We used a 5x161 point $x$-$z$ grid, with periodic
boundary conditions in the $x$ direction.  We use the temperature and
tangential velocity boundary condition for smooth surfaces found in
reference \cite{jenkins92}.  We can alternate between our new
\eqref{eq:nbc} and the standard $\Vect{v}\cdot\hat{n}=0$ boundary
conditions for the normal velocity.  To maintain mass conservation we
use a special differencing procedure for the mass continuity equation,
which is second-order accurate for both the new and old boundary
conditions and is exactly conserving.  In one dimension we label the
density and velocity $\rho_i$ and $w_i$ for $i=0$ to $N$.  We then
define $N+2$ fluxes $\Phi_i=(\rho_{i-1}+\rho_i)(w_{i-1}+w_i)/4$ for
$i=1$ to $N$, and $\Phi_0=\Phi_{N+1}=0$.  Then the change in density
at position $z_i=i\Delta z$ is $\Delta\rho_i=(\Phi_{i+1}-\Phi_i)\Delta
t$, where $\Delta t$ is the current time step.  From these definitions
it follows that $\sum_{i=0}^N \Delta\rho_i=0$ and $\sum_{i=0}^N
\rho_i=\text{constant}$, regardless of the value of $w$ and $\rho$ at
the boundaries.  The only boundaries in the system are the actual
walls.  We let the free surfaces develop on there own simply as very
low density regions.  As the density of a gas becomes so dilute that
the mean free path is comparable to the size of the container the
transport equations must be modified.  We define a transport cutoff at
a density of
\eq
\nu_0=\frac{1}{6l_0\sqrt{2}}=5.89 x 10^{-3},
\eeq
where $l_0=20$ is the dimensionless cutoff mean free path.  At lower
densities than this the viscosity and thermal conductivity are
multiplied by the factor 
\eq
\chi=\frac{\nu^2}{\nu_0^2+\nu^2}.
\eeq
For $\nu>>\nu_0$ $\chi$ is close to unity, but as the density
decreases the transport coefficients go smoothly to zero.  This ad-hoc
factor is used only to prevent the code from diverging at extremely
low densities. In these simulations, we have varied from 10-100 and
the results are almost the same except for slight deviations in the
very low-density regions.  This low density regime does not effect the
main flow as the momentum and energy are also nearly zero unless the
temperature or velocity diverge, which is prevented by this approach.

We use the following parameter for all of the simulations presented
here: $e_0=0.7$ and $T0=1$.  Both of these parameter effect the energy
loss.  If the energy loss is not great enough the separation does not
occur.  These values were not formally optimized, but several
different value of $e_0$ between 0.6 and 0.9 were tried, but anything
below 0.8 gave approximately the same flight-time for the vibrated
case.  The temperature and tangential velocity boundary conditions
require two additional parameters, $e_w=.5$ and $\mu_w=0$, the wall
coefficients of restitution and friction.  These parameters only have
a small effect on the flow.  

For the free-fall case we solve the equation in the lab frame.  To
create an initial condition we set gravity upwards to push the fluid
to the top of the container.  Then we slowly decrease gravity to zero
and then abruptly change the value of $g$ to $-g$. The results of the
horizontally averaged density are shown as space-time plots in
\figref{spacetime} for both the new and old boundary conditions.  For
the vibrated case we solve the problem in the reference frame of the
bottom plate.  We determine the acceleration of the plate from the
experimentally determined position of the plate \figref{plate}.  The
second derivative is quite noisy so the signal is convolved with a 5
ms top-hat function.  The signal is then corrected so that the average
acceleration is zero.  The initial condition is a uniform distribution
and wait for the density to reach a stable periodic state.  This
typically takes 5-10 oscillations.  The final cycle is shown in
\figref{spacetime} for both types of boundary conditions.

\section{Results}
Comparing the first two columns in \figref{spacetime} (experiment and
new boundary conditions) to the third there is a striking effect.  In
the free-fall experiment the zero velocity boundary condition holds
the fluid to the top surface long after all of the grains in the
experiment and the new boundary condition simulation have left the
domain of interest.  To see this in a quantitative way we look at the
rate the particles pass the lower line in the figure.  The result is
shown in \subfigref{compare}{a}.  Initially the two simulation follow
more closely due to the lack of dispersion in the experiment.  The
dispersion in the simulation is not due to temperature, but an
artifact of the truncation error in the mass conservation equations.
Shortly ($~20$ ms) all three curves meet, but by ($~30$ ms) the old
boundary conditions simulation begins to diverge, and from there on
the behavior is radically different.  As the last of the material
flows from the experiment and new BC simulation there is another small
lag, but the old BC simulation continues to flow for another 30 ms.
\Subfigref{compare}{c} shows that there is a negative acceleration induced
by the old BCs.  Both the experiment and the new BCs show zero
acceleration throughout the time of the experiment.

In \subfigref{spacetime}{b} the layer is stuck to the boundary for the
old BCs.  There is no clear flight time.  From the layer acceleration
$a_b$ in \subfigref{compare}{b} we see that there is still some kind
of a collision for the old BCs but it occurs earlier.  Further, there
is a clear region of negative acceleration, as seen in the expanded
axis of \subfigref{compare}{d} of nearly half the value of gravity.
The small negative acceleration in the experiment is likely due to
friction with the walls.  The take off time and collision time shown
by the two vertical line are the same for the experiment and the new
BCs.

\section{Conclusions}

The new boundary conditions provide a significant improvement in
simulating these flows.  Separation (cavitation) is a very common
situation in granular flows, and this type of boundary condition is
needed to capture the basic features of the flows presented.  Without
this type of boundary condition many qualitative features may be
missed, and there is no hope quantitative agreement. There are still a
number of open question including: what should happen in corners, and
how can these BCs be treated analytically?  Previously we have shown
that molecular dynamics (MD) simulation \cite{bizon98,moon2002} are
capable of quantitatively reproducing the flow in dense vibrated
pattern forming system, and MD and continuum simulation can reproduce
dilute flows in which granular shocks form \cite{rericha2002}.
However, this work represents a first, big step in getting quantitative
agreement using continuum simulations in dense vibrated pattern
forming systems.

\begin{acknowledgments}
This work is supported by The National Science Foundation, Math,
Physical Sciences Department of Materials Research under the Faculty
Early Career Development (CAREER) Program: DMR-0134837.
\end{acknowledgments}


\bibliography{\bd/sand,\bd/shattuck,local}

\end{document}